\documentclass[oldversion]{aa}

\usepackage{epsfig}
\usepackage{graphics}
\usepackage{float}
\usepackage{amsmath}
\usepackage{multirow}
\usepackage{longtable}
\usepackage{rotate}
\usepackage{array}
\usepackage{subfigure}
\def\kms{\ifmmode{\rm km\,s^{-1}}\else\hbox{$\rm km\,s^{-1}$}\fi}

\setlongtables

\begin{document}

\title{Frequentist tests for Bayesian models}

\author{L.B.Lucy}
\offprints{L.B.Lucy}
\institute{Astrophysics Group, Blackett Laboratory, Imperial College 
London, Prince Consort Road, London SW7 2AZ, UK}
\date{Received ; Accepted }

\abstract{Analogues of the frequentist chi-square and $F$ tests are 
proposed for testing goodness-of-fit and consistency for Bayesian models.
Simple examples exhibit these tests' detection of 
inconsistency between consecutive experiments with identical parameters, 
when the first experiment 
provides the prior for the second. In a related analysis, 
a quantitative measure is derived for judging the degree of tension
between two different experiments with partially overlapping 
parameter vectors. 
}
\authorrunning{Lucy}
\titlerunning{Tests for Bayesian models}
\maketitle
%________________________________________________________________
%
%
%
%
%
\section{Introduction}
Bayesian statistical methods are now widely applied in astronomy.
Of the new techniques thus introduced, model selection (or comparison)
is especially notable
in that it replaces the frequentist acceptance-rejection 
paradigm for testing hypotheses. Thus, given a data set $D$, there 
might be several hypotheses $\{H_{k}\}$ that have 
the potential to
explain $D$. From these, Bayesian model selection identifies the
particular $H_{k}$ that best explains $D$. The procedure is simple, though
computationally demanding: starting with an arbitrary pair of the 
$\{H_{k}\}$, we apply the model selection machinery, discard the weaker 
hypothesis, replace it with one of the remaining ${H_{k}}$, and repeat. 

This procedure usefully disposes of the weakest hypotheses, but
there is no guarantee that the surviving $H_{k}$ explains
$D$. If the correct hypothesis is not included 
in the $\{H_{k}\}$, we are left with the 'best'
hypothesis but are not made aware that the search for an explanation
should continue. In the context of model selection, the next step would
be a comparison of this 'best' $H_{k}$ with the hypothesis that $H_{k}$  
is false. But model selection fails at this point because we cannot compute
the required likelihood (Sivia \& Skilling 2006, p.84).
Clearly, what is needed is a 
goodness-of-fit criterion for Bayesian models.         

This issue is discussed by Press et al. (2007, p.779). They note that
``There are no good fully Bayesian methods for assessing goodness-of-fit
...'' and go on to report that ``Sensible Bayesians usually fall back to 
$p$-value tail statistics ...when they really need to know if a model is 
wrong.''

On the assumption that astronomers do
really need to know if their models are wrong, this paper adopts
a frequentest approach to testing Bayesian models.
Although this approach may be immediately abhorrent to committed Bayesians,
the role of the tests proposed herein is merely to provide a 
quantitative
measure according to which Bayesians decide whether their models are
satisfactory. When they are, the Bayesian inferences are presented - and
with increased
confidence. When not, flaws in the models or the data must be sought,
with the aim of eventually achieving satisfactory Bayesian inferences.
\section{Bayesian models}
The term Bayesian model - subsequently denoted by ${\cal M}$ - must now be 
defined. The natural
definition of ${\cal M}$ is that which must be specified in order that 
Bayesian inferences can be drawn from $D$ - i.e., in order to compute  
posterior probabilities.
This definition implies that, 
in addition to the hypothesis $H$, which introduces the parameter vector
$\vec{\alpha}$,
the prior probability distribution $\pi(\vec{\alpha})$
must also be included in ${\cal M}$. Thus, symbolically, we write
\begin{equation}
  {\cal M} \equiv  \left\{\pi,H \right\}  
\end{equation}
It follows that different Bayesian models
can share a common $H$. For example, $H$ may be the hypothesis
that $D$ is due to Keplerian motion. 
But for the motion of a star about the Galaxy's central black hole,
the appropriate $\pi$ will differ from that for the reflex orbit of 
star due to a planetary companion.

A further consequence is that a failure of ${\cal M}$
to explain $D$ is not necessarily due to $H$: an inappropriate
$\pi$ is also a possibility.

To astronomers accustomed to working only with uniform
priors, the notion that a Bayesian model's poor fit to $D$ could be due to 
$\pi$ might be surprising. A specific
circumstance where $\pi$ could be at fault arises when Bayesian
methods are used to repeatedly update our knowledge of some phenomenon -
e.g., the value of a fundamental  
constant that over the years is the subject of 
numerous experiments ($X_{1}, \dots,X_{i},\dots$), usually of 
increasing precision.
With an orthodox Bayesian approach, $X_{i+1}$ is analysed with the prior 
set equal to the posterior from $X_{i}$. Thus
\begin{equation}
  \pi_{i+1}(\vec{\alpha}) = p(\vec{\alpha}|H,D_{i}) 
\end{equation}
This is the classical use of Bayes theorem to update our opinion by
incorporating past 
experience. However, if 
$X_{i}$ is flawed - e.g., due to an unrecognized
systematic error - then this choice of $\pi$ impacts negatively on the
analysis of $X_{i+1}$, leading perhaps to a poor fit to $D_{i+1}$ 

Now, since subsequent flawless experiments result in the decay of
the negative impact of $X_{i}$, this recursive procedure is self-correcting,
so a Bayesian might argue that the problem can be ignored.
But scientists feel obliged to resolve such discrepancies before
publishing or undertaking further experiments and therefore need a
statistic that quantifies any failure of ${\cal M}$ to explain $D$.
\section{A goodness-of-fit statistic for Bayesian models}
The most widely used goodness-of-fit statistic in the frequentist
approach to hypothesis testing
is $\chi^{2}$, whose value is determined by the residuals
between the fitted model and the data, with no input from prior knowledge.
Thus,
\begin{equation}
 \chi^{2}_{0} = \chi^{2}(\vec{\alpha}_{0})
\end{equation}
is the goodnes-of-fit statistic for the minimum-$\chi^{2}$ solution
$\vec{\alpha}_{0}$.

A simple analogue of $\chi^{2}_{0}$ for Bayesian models is the posterior 
mean of $\chi^{2}(\vec{\alpha})$,
\begin{equation}
  \langle \chi^{2} \rangle_{\pi} \: = \int \chi^{2}(\vec{\alpha}) \: 
                                     p(\vec{\alpha}|H,D) \: dV_{\vec{\alpha}}
\end{equation}
where the posterior distribution 
\begin{equation}
   p(\vec{\alpha}|H,D)     = 
 \frac{ \pi (\vec{\alpha}) {\cal L}(\vec{\alpha}|H,D)}
              {\int \pi (\vec{\alpha}) 
              {\cal L}(\vec{\alpha}|H,D) \: dV_{\vec{\alpha}}} 
\end{equation}
Here ${\cal L}(\vec{\alpha}|H,D)$ is the likelihood of 
$\vec{\alpha}$ given data $D$.

Note that since $\langle \chi^{2} \rangle_{\pi}$ depends on both constituents of ${\cal M}$, 
namely $H$ {\em and} $\pi$, it has the potential to detect a poor fit due 
to either or
both being at fault, as required by Sect.2.  

In Eq.(4) the subscript $\pi$ is attached to $\langle\chi^{2}\rangle$ to 
stress that
a non-trivial, informative prior is included in ${\cal M}$. On the other
hand, when a uniform prior is assumed, $\langle\chi^{2}\rangle$ is 
independent of
the prior and is then denoted by $\langle\chi^{2}\rangle_{u}$.

The Bayesian goodness-of-fit statistic $\langle\chi^{2}\rangle_{u}$ is used in 
Lucy (2014; L14) to illustrate
the detection of a weak second orbit in simulations of {\em Gaia} scans
of an astrometric binary. In that case, $H$ states that the
scan residuals are due to the reflex Keplerian orbit caused by {\em one}
invisible companion. With increasing amplitude of the second orbit, the point
comes when the investigator will surely abandon $H$ - i.e., abandon the 
assumption of just one companion - see Fig.12, L14.
\subsection{$P$-values}
With the classical frequentist acceptance-rejection paradigm,
a null hypothesis $H_{0}$ is rejected on the basis of a $p$-value tail 
statistic.
Thus, with the 
$\chi^{2}$ test, $H_{0}$ is rejected if $\chi^{2}_{0} > \chi^{2}_{\nu,\beta}$, 
where $p(\chi^{2}_{\nu} > \chi^{2}_{\nu,\beta}) = \beta$, and accepted 
otherwise.
Here $\nu = n - k$ is the number of degrees of freedom, where $n$ is the 
number of measurements and $k$ is the
number of parameters introduced by $H_{0}$, and $\beta$ is the designated
$p$-threshold chosen by the investigator. 

For a Bayesian model, a $p$-value can be computed from 
the $\langle\chi^{2}\rangle_{\pi}$ statistic, whose approximate distribution
is derived below in Sect.5.1. However, a sharp transition from acceptance to
rejection of the null hypothesis at some designated $p$-value is not
recommended. First, $p$-values overstate the strength of the evidence 
against $H_{0}$ (e.g., Sellke et al. 2001). In particular, the value 
$p = 0.05$ recommended in elementary texts does not imply strong evidence 
againts $H_{0}$.
Second, the $p$-value is best regarded (Sivia \& Skilling 2006, p.85)   
as serving a qualitative purpose, with a small value prompting us to
think about alternative hypotheses.  
Thus, if $\langle \chi^{2} \rangle _{\pi}$ exceeds
the chosen threshold $\chi^{2}_{\nu,\beta}$, this is a warning that something is 
amiss and
should be investigated, with the degree of concern increasing as $\beta$
decreases. If the $\beta = 0.001$ threshold is exceeded, then the
investigator would be well-advised to suspect that ${\cal M}$ or $D$ 
is at fault.    

Although statistics texts emphasize tests of $H$ not $D$, 
astronomers know that $D$ can be corrupted by biases or calibration 
errors. Departures from normally-distributed errors can also increase 
$\langle\chi^{2}\rangle_{\pi}$.

If $D$ is not at fault, then ${\cal M}$ is the culprit, implying that
either $\pi$ or $H$ is at fault. If the fault lies with $\pi$ not
$H$ , then
we expect that $\langle\chi^{2}\rangle_{u} \: < \: \chi^{2}_{\nu,\beta}$ even 
though
\newline $\langle \chi^{2} \rangle _{\pi} \: > \: \chi^{2}_{\nu,\beta}$.

If neither $D$ nor $\pi$ can be faulted, then the investigator must
seek a refined or alternative $H$. 
\subsection{Type I and type II errors}
In the frequentist approach to hypothesis testing, decision errors
are said to be of type I if $H$ is true but the test says reject $H$, and of
type II if $H$ is false but the test says accept $H$.

Since testing a Bayesian model is not concerned exclusively  
with $H$,
these definitions must be revised, as follows:\\
A type I error arises
when ${\cal M}$ and $D$ are flawless but the statistic
(e.g., $\langle \chi^{2} \rangle_{\pi}$) exceeds the designated threshold.\\
A type II error arise when ${\cal M}$ or $D$ are flawed but the statistic
does not exceed the designated threshold.\\

Here the words accept and reject are avoided. Moreover, no particular 
threshold is mandatory: it is at the discretion of the investigator and 
is chosen with regard to the consequences of making a decision error. 
\section{Statistics of $\langle \chi^{2} \rangle$}
The intuitive understanding that scientists have regarding 
$\chi^{2}_{0}$ derives from its simplicity and the 
rigorous theorems on its distribution that allow us
to derive confidence regions for multi-dimensional linear models
(e.g., Press et al. 2007, Sect.15.6.4). 

Rigorous statistics for $\chi^{2}$ require two assumptions: 1) that the 
model fitted to $D$
is linear in its parameters, and 2) that measurement errors obey
the normal distribution. Nevertheless,
even when these standard assumptions do not strictly
hold, scientists still commonly rely on $\chi^{2}_{0}$ to gauge 
goodness-of-fit,
with perhaps Monte Carlo (MC) sampling to provide justification 
or calibration (e.g., Press et al. 2007, Sect.15.6.1).

Rigorous results for the statistic $\langle\chi^{2}\rangle$ are therefore
of interest. In fact,
if we add the assumption of a uniform prior to the above standard 
assumptions, then we may prove 
(Appendix A) that   
\begin{equation}
  \langle \chi^{2} \rangle_{u} \: =  \chi^{2}_{0} + k
\end{equation}
where $k$ is the number of parameters.

Given that Eq.(6) is exact under the stated assumptions, it follows that
the quantity  $\langle \chi^{2} \rangle_{u} - k$ is distributed as
$\chi^{2}_{\nu}$ with $\nu = n-k$ degrees of freedom.

For minimum-$\chi^{2}$ fitting of linear models, the solution is always a 
better fit to $D$ than is the true solution. In particular, 
${\cal E}(\chi^{2}_{true}) = n$, whereas  
${\cal E}(\chi^{2}_{0}) = n - k$.
Accordingly, the $+k$ term in Eq.(6) 'corrects' $\chi^{2}_{0}$ for
overfitting and so  
${\cal E} (\langle \chi^{2} \rangle_{u}) = {\cal E} (\chi^{2}_{true})$ 
- i.e., the expected value of $\chi^{2}$ for 
the actual measurement errors. 
\subsection{Effect of an informative prior}
When an informative prior is included in ${\cal M}$, an analytic formula
for $\langle \chi^{2} \rangle_{\pi}$ is in general not available. However,
its approximate statistical properties are readily found. 

Consider again a linear model with normally-distributed errors and suppose
further that the experiment ($X_{2}$) is without flaws. The $\chi^{2}$
surfaces   
are then self-similar $k$-dimensional ellipsoids with minimum at 
$\vec{\alpha_{0}} \approx \vec{\alpha_{true}}$, 
the unknown exact solution. Let us now further suppose that
the informative prior $\pi(\vec{\alpha})$ derives from a previous flawless
experiment ($X_{1}$). The prior $\pi$ will then be a convex 
(bell-shaped) function with maximum at 
$\vec{\alpha}_{max} \approx  \vec{\alpha}_{0}$.
Now, consider a 1-D family of such functions all centred on
$\vec{\alpha}_{0}$ and ranging from the very broad to the very narrow. For the 
former 
$\langle \chi^{2} \rangle_{\pi} \: \approx \: \langle \chi^{2} \rangle_{u}$; 
for the latter
$\langle \chi^{2} \rangle_{\pi} \: \approx \: \chi^{2}_{0}$. Thus, ideally,
when a Bayesian model ${\cal M}$ is applied to data $D$ 
we expect that
\begin{equation}
      \langle\chi^{2} \rangle_{u}  \;\;  \ga \;\; \langle\chi^{2} \rangle_{\pi} \;\; 
                                           \geq \;\;  \chi^{2}_{0} 
\end{equation}

Now, a uniform prior and $\delta(\vec{\alpha} - \vec{\alpha_{0}})$ are
the limits of the above family of bell-shaped functions. Since the delta 
function 
limit is not likely to be closely approached in practice, a first
approximation to the distribution of  $\langle\chi^{2} \rangle_{\pi}$ is
that of  $\langle\chi^{2} \rangle_{u}$ - i.e., that of $\chi^{2}_{n-k} + k$. 

The above discussion assumes faultless $X_{1}$ and $X_{2}$.
But now suppose that there is an inconsistency between $\pi$ and $X_{2}$.
The peak of $\pi$ at $\vec{\alpha}_{max}$ will then in general be
offset 
from the minimum of $\chi^{2}$ at $\vec{\alpha}_{0}$. Accordingly,
in the calculation of $\langle\chi^{2} \rangle_{\pi}$ from Eq.(4),
the neighbourhood of the $\chi^{2}$ minimum $\chi^{2}_{0}$ at 
$\vec{\alpha}_{0}$ has reduced weight relative to 
$\chi^{2}(\vec{\alpha}_{max})$ at the peak of $\pi$. Evidently, in this 
circumstance, $\langle\chi^{2} \rangle_{\pi}$ can greatly exceed 
$\langle \chi^{2} \rangle_{u}$,
and the investigator is then alerted to the inconsistency.
\section{Numerical experiments}
Given that rigorous results $\langle \chi^{2} \rangle$ are not available for 
informative priors, numerical tests are essential to illustrate the 
discussion of Sect.4.1.
\subsection{A toy model}
A simple example with just one parameter $\mu$ is as follows:
$H$ states that $u=\mu$, and $D$ comprises $n$ 
measurements
$u_{i} = \mu + \sigma z_{i}$, where the $z_{i}$ here and below are independent 
gaussian variates randomly sampling ${\cal N}(0,1)$. In creating synthetic data,
we set $\mu = 0, \sigma = 1$ and $n = 100$.

In the first numerical experiment, two independent data sets 
$D_{1}$ and 
$D_{2}$ are created comprising $n_{1}$ and $n_{2}$ measurements, respectively. 
On the assumption of a uniform prior, the posterior density of $\mu$ 
derived from $D_{1}$ is
\begin{equation}
  p(\mu|H,D_{1}) = \frac{{\cal L}(\mu|H,D_{1})} 
                      {\int {\cal L}(\mu|H,D_{1}) \: d\mu}
\end{equation}
We now regard $p(\mu|H,D_{1})$ as prior knowledge to be taken into account 
in analysing $D_{2}$. Thus
\begin{equation}
 \pi(\mu) =  p(\mu|H,D_{1}) 
\end{equation}
so that the posterior distribution derived from $D_{2}$ is
\begin{equation}
  p(\mu|H,D_{2}) = \frac{\pi(\mu) {\cal L}(\mu|H,D_{2})} 
              {\int {\pi(\mu) \cal L}(\mu|H,D_{2}) \:d\mu}
\end{equation}
The statistic quantifying the goodness-of-fit achieved when
${\cal M} = \{\pi,H\}$ is applied to data $D_{2}$ is then 
\begin{equation}
  \langle \chi^{2} \rangle_{\pi} \: = \: \int \chi^{2}(\mu) \: p(\mu|H,D_{2}) \: d\mu
\end{equation}

From $N$ independent data pairs $(D_{1},D_{2})$, we obtain $N$ independent
values of  $\langle \chi^{2} \rangle_{\pi}$ thus allowing us to test the expectation 
(Sect.4.1) that this statistic is approximately distributed as 
$\langle \chi^{2} \rangle_{u}$. 
In Fig.1, this empirical PDF is plotted together the theoretical PDF
for  $\langle \chi^{2} \rangle_{u}$. 
\begin{figure}
\vspace{8.2cm}
\includegraphics{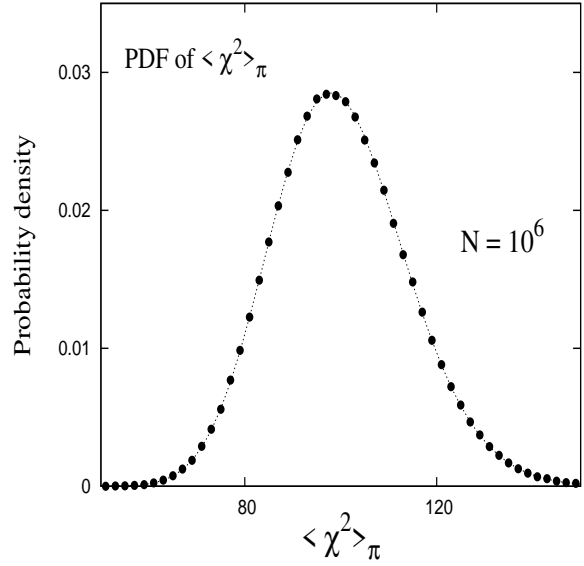}
\caption{Empirical PDF of $\langle \chi^{2} \rangle_{\pi}$ derived from 
the analysis of $10^{6}$ data pairs $D_{1},D_{2}$ as described in Sect.5.1. 
The dashed curve is the theoretical PDF for $\langle\chi^{2}\rangle_{u}$.} 
\end{figure}
The accuracy of the approximation at large values of $\chi^{2}$ is
of special importance. For $n=100$ and $k=1$, the 0.05,0.01 and 0.001
critical points of $\langle\chi^{2}\rangle_{u}$ are 124.2, 135.6 and 149.2,
respectively. From a simulation with $N = 10^{6}$, the number of
$\langle\chi^{2}\rangle_{\pi}$ values exceeding
these thresholds are 50177, 10011 and 1025, respectively. Thus, the
fraction of $\langle\chi^{2}\rangle_{\pi}$ exceeding the critical values derived
from the distribution of $\langle\chi^{2}\rangle_{u}$ are close to their 
predicted values.
Accordingly, the conventional interpretation of these critical values is
valid.

In this experiment, the analysis of $X_{2}$ benefits from knowledge gained
from $X_{1}$. We expect therefore that  
$\langle\chi^{2}\rangle_{\pi}$ is less than $\langle\chi^{2}\rangle_{u}$, since 
replacing the uniform
prior with the informative $\pi$ obtained from $X_{1}$ should improve the
fit. From $10^{6}$ repetitions, this proves to be so with probability
0.683. Sampling noise in $D_{1}$ and $D_{2}$ accounts for the shortfall.
\subsection{Bias test}
When $X_{1}$ and $X_{2}$ are flawless, the 
statistic 
$\langle \chi^{2}\rangle_{\pi}\!$ indicates doubts - i.e., type I errors 
(Sect.3.2) - 
about the mutual consistency of $X_{1}$ and $X_{2}$
with just the expected frequency. Thus, with the $5\%$ 
threshold, doubts arise in $5.02 \%$ of the above $10^{6}$ trials. This 
encourages the use of $\langle\chi^{2} \rangle_{\pi}$ to detect inconsistency.

Accordingly, in a second test, $X_{1}$ is flawed due to biased 
measurements.
Thus, now $u_{i} = \mu + \sigma z_{i} + b$ for $D_{1}$.
As a result,
the prior for $X_{2}$ obtained from Eq.(9) is compromised, and this  
impacts on the statistic $\langle\chi^{2} \rangle_{\pi}$ from Eq.(11).

In Fig.2, the values of $\langle\chi^{2} \rangle_{\pi}$ and 
$\langle\chi^{2} \rangle_{u}$
are plotted against $b/\sigma$. Since the compromised prior is excluded
from $\langle\chi^{2} \rangle_{u}$, its values depend only on the flawless 
data 
sets $D_{2}$, and so mostly fall within the $(0.05,0.95)$ interval.  
In contrast, as $b/\sigma$ increases, the values of 
$\langle\chi^{2} \rangle_{\pi}$ are
increasingly affected by the compromised prior. 

The mutual consistency of $X_{1}$ and $X_{2}$ is assessed
on the basis of $\langle\chi^{2} \rangle_{\pi}\:$, with choice of critical
level at our discretion. However, when
$\langle\chi^{2} \rangle_{\pi}$ exceeds the $0.1\%$ level at $149.2$, we
would surely conclude that $X_{1}$ and $X_{2}$ are in conflict and 
seek to resolve the discrepancy. On the other hand, when inconsistency is not
indicated, we may accept the Bayesian inferences
derived from $X_{2}$
in the confident belief that incorporating prior knowledge from $X_{1}$ is
justified and beneficial.
\begin{figure}
\vspace{8.2cm}
\includegraphics{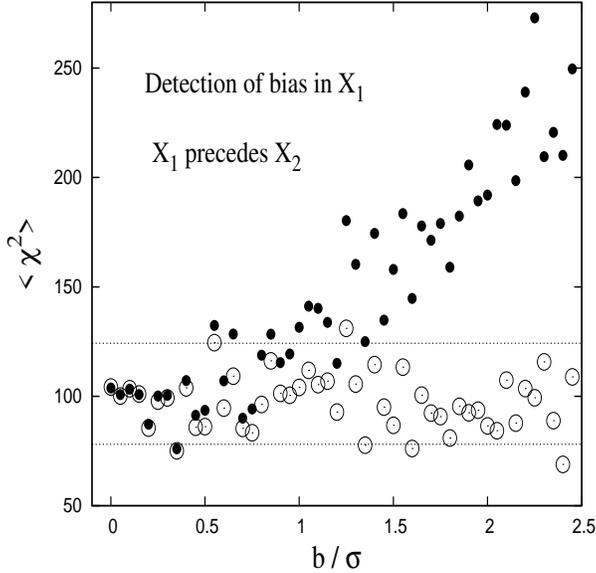}
\caption{Detection of bias in $X_{1}$ with the $\langle\chi^{2}\rangle_{\pi}$ 
statistic when $X_{2}$ is analysed with prior derived from $X_{1}$.
Values of $\langle\chi^{2}\rangle_{\pi}$ (filled circles)
and $\langle\chi^{2}\rangle_{u}$ (open circles) are plotted against the bias 
parameter $b/\sigma$.
The dashed lines are the $5$ and $95\%$ levels.} 
\end{figure}
This test illustrates the 
important point that an inappropriate $\pi$ can corrupt the 
Bayesian inferences drawn from a flawless experiment. Thus, in this case,
the bias in $D_{1}$ propagates into the posterior $p(\mu| H,D_{2})$
derived from $X_{2}$.
This can be (and is) avoided by preferring the frequentist methodology.
But to do so is to forgo the great merit of Bayesian inference,
namely its ability to incorporate {\em informative} prior information
(Feldman \& Cousins 1998). If one does therefore prefer Bayesian 
inference, it is evident that a goodness-of-fit statistic such as 
$\langle\chi^{2}\rangle_{\pi}$ is 
essential in order to detect errors propagating into the posterior
distribution from an ill-chosen prior.
\subsection{Order reversed}
In the above test, the analysis of $X_{2}$ is preceded by that
of $X_{1}$. This order can be reversed. Thus, with the {\em same} $N$ data
pairs ($D_{1},D_{2}$), we now first analyse $X_{2}$ with a uniform prior
to obtain $p(\mu|H,D_{2})$. This becomes the prior for the analysis
of $X_{1}$. This analysis then gives the posterior $p(\mu|H,D_{1})$ from which
a new value of $\langle\chi^{2}\rangle_{\pi}$ is obtained.

When the values of $\langle\chi^{2}\rangle_{\pi}$ obtained with this
reversed order of analysis are plotted against $b/\sigma$, the result
is similar to Fig.2, implying that the order is unimportant.
Indeed, statistically, the same decision is reached independently of order.
For example, for $10^{5}$ independent data pairs $(D_{1},D_{2})$ with
$b/\sigma = 1$, the number with 
$\langle\chi^{2}\rangle_{\pi} > 124.2$, the $5\%$ threshold, is 
$50267$ when $X_{1}$ precedes $X_{2}$ and
$50149$ when $X_{2}$ precedes $X_{1}$.

\subsection{Alternative statistic}

Noting that the Bayesian evidence is $= \bar{ {\cal L}}$, the prior-weighted 
mean of the likelihood, we can, under standard assumptions, write
\begin{equation}
    \bar{ {\cal L}} \propto \exp\left[-\frac{1}{2} \chi^{2}_{\rm eff} \right] 
\end{equation}
where the effective $\chi^{2}$ (Bridges et al. 2009) is
\begin{equation}
   \chi^{2}_{\rm eff} \: = \: - 2 \ln \int \pi(\vec{\alpha}) 
          \exp\left[-\frac{1}{2} \chi^{2}(\vec{\alpha}) \right] dV_{\vec{\alpha}} 
\end{equation}
This is a possible alternative to $\langle \chi^{2}  \rangle_{\pi}$ defined
in Eq.(4). However, in the test of Sect.5.1, the two values are so nearly
identical it is immaterial which mean is used. Here 
$\langle \chi^{2}  \rangle_{\pi}$ is preferred because it remains well-defined
for a uniform prior, for which an analytic result is available (Appendix A).

Because $\langle \chi^{2}  \rangle_{\pi}$ and $\chi^{2}_{\rm eff}$ are nearly 
identical, the distribution of $\chi^{2}_{\rm eff}$  should approximate that of
$\langle \chi^{2}  \rangle_{u}$ (Sect.4.1). To test this, the experiment of
Sect.5.1 is repeated with  $\chi^{2}_{\rm eff}$ replacing 
$\langle \chi^{2}  \rangle_{\pi}$. From a simulation with $N = 10^{6}$,
the number of  $\chi^{2}_{\rm eff}$ values exceeding the 0.05,0.01 and 0.001 
thresholds
are 50167, 9951 and 970, respectively. Thus if  $\chi^{2}_{\rm eff}$ is chosen
as the goodness-of-fit statistic, accurate $p$-values can be derived on 
the assumption that  $\chi^{2}_{\rm eff}-k$ is distributed as $\chi^{2}_{\nu}$ 
with
$\nu = n-k$ degrees of freedom. From Sect.4.1, we expect these $p$-values to
be accurate if $\pi(\vec{\alpha})$ is {\em not} more sharply peaked than
${\cal L}(\vec{\alpha}|H,D)$.

\section{An $F$ statistic for Bayesian models}
Inspection of Fig.2 shows that a more powerful test
of inconsistency between $X_{1}$ and $X_{2}$ must exist. A systematic 
displacement of $\langle\chi^{2}\rangle_{\pi}$ relative
to $\langle\chi^{2}\rangle_{u}$ is already evident even when  
$\langle\chi^{2}\rangle_{\pi}$ is below the $5\%$ threshold
at 124.2. This suggests that a Bayesian analogue of the $F$ statistic be
constructed.

\subsection{The frequentist $F$-test}

In frequentist statistics, a standard result (e.g., Hamilton 1964, p.139) 
in the testing of {\em linear} hypotheses
is the following: we define the statistic 
\begin{equation}
   F = \frac{n-i}{j} \: \frac{\chi^{2}_{c} - \chi^{2}_{0}}{\chi^{2}_{0}}
\end{equation}
where $\chi^{2}_{0}$ is the minimum value of
$\chi^{2}$ when all $i$ parameters are adjusted, and $\chi^{2}_{c}$ is
the minimum value when a linear constraint is imposed on $j \: (\leq i)$ 
parameters, so that only the remaining $i-j$ are adjusted. Then, on 
the null hypothesis $H$ 
that the constraint is true, $F$ is distributed as $F_{\nu_{1},\nu_{2}}$
with $\nu_{1} = j$ and $\nu_{2} = n-i$, where
$n$ is the number of measurements. Accordingly, $H$ 
is tested by comparing the value
$F$ given by Eq.(14) with critical values $F_{\nu_{1},\nu_{2},\beta}$ derived 
from the distribution of $F_{\nu_{1},\nu_{2}}$. 

Note that when 
$j = i$, the constraint completely determines $\vec{\alpha}$. If this
value is $\vec{\alpha}_{*}$, then
$\chi^{2}_{c} = \chi^{2}(\vec{\alpha}_{*})$ and $H$ states that
$\vec{\alpha}_{*} = \vec{\alpha}_{true}$.
 
A particular merit of the statistic $F$ is that it is
independent of $\sigma$. However, the resulting $F$-test 
does assume normally-distributed measurement errors.

\subsection{A Bayesian $F$}
   
In a Bayesian context, the frequentist hypothesis that  
$\vec{\alpha}_{true} = \vec{\alpha}_{*}$ is replaced by 
the statement that $\vec{\alpha}_{true}$ obeys  
the posterior distribution $p(\vec{\alpha}|H,D_{2})$. 
Thus an exact constraint is replaced by a fuzzy constaint.

Adopting the simplest approach, we define ${\cal F}$, a Bayesian 
analogue of $F$,
by taking  
$\chi^{2}_{c}$ to
be the value at the posterior mean of $\vec{\alpha}$, 
\begin{equation}
 \langle \vec{\alpha} \rangle_{\pi} \: = \: 
  \frac{\int \vec{\alpha} \: \pi( \vec{\alpha} ){\cal L}(\vec{\alpha}|H,D_{2})   \: dV_{\vec{\alpha}}}
  {\int  \pi( \vec{\alpha} ){\cal L}(\vec{\alpha}|H,D_{2} ) \: dV_{\vec{\alpha}}} 
\end{equation}
where $\pi(\vec{\alpha})$ is the informative prior.
Considerations of accuracy when values of $\chi^{2}$ are computed on
a grid suggest we take $\chi^{2}_{0}$ to be the value at  
\begin{equation}
 \langle \vec{\alpha} \rangle_{u} \: = \: 
  \frac{\int \vec{\alpha} \: {\cal L}(\vec{\alpha}|H,D_{2})   \: dV_{\vec{\alpha}}}
  {\int {\cal L}(\vec{\alpha}|H,D_{2} ) \: dV_{\vec{\alpha}}} 
\end{equation}
the posterior mean for a uniform prior.

With  $\chi^{2}_{c}$ and $\chi^{2}_{o}$ thus defined, the 
Bayesian $F$ is 
\begin{equation}
 {\cal F} = \frac{n-i}{j} \: \frac{\chi^{2}(\langle \vec{\alpha} \rangle_{\pi}) 
    - \chi^{2}(\langle \vec{\alpha} \rangle_{u})}
       { \chi^{2}(\langle \vec{\alpha} \rangle_{u})}
\end{equation}
and its value is to be compared with
the chosen threshold $F_{\nu_{1},\nu_{2},\beta}$ when
testing the consistency of $X_{1}$ and  $X_{2}$. Since 
$\langle \vec{\alpha} \rangle_{u}$ is independent of $\pi(\vec{\alpha})$,
the statistic ${\cal F}$ measures the effect of $\pi(\vec{\alpha})$
in displacing $\langle \vec{\alpha} \rangle_{\pi}$ from 
$\langle \vec{\alpha} \rangle_{u}$.

In this Bayesian version of the $F$-test, the null hypothesis states
that the posterior mean
$\langle \vec{\alpha} \rangle_{\pi}  = \vec{\alpha}_{true}$. This will
be approximately true 
when a flawless Bayesian model ${\cal M}$ is applied to a flawless
data set $D$. However, if the chosen threshold 
on ${\cal F}$ is exceeded, then one or more of $\pi,H$ and $D$ is 
suspect as discussed in Sect.3.1. If the threshold is not exceeded, then
Bayesian inferences drawn from the posterior distribution
$p(\vec{\alpha}|H,D_{2})$ are supported.

\subsection{Test of $p$-values from ${\cal F}$}

If Eq.(17) gives ${\cal F} = {\cal F}_{*}$, the corresponding $p$-value
is
\begin{equation}
  p_{*} = \int_{ {\cal F}_{*}}^{\infty} \: F_{\nu_{1},\nu_{2}} \: dF
\end{equation}
where $\nu_{1} = j$ and  $\nu_{2} = n-i$.
The accuracy of these $p$-values can be tested with the 1-D toy model
of Sect.5.1 as follows:\\

(i) Independent data sets $D_{1},D_{2}$ are created corresponding to 
    $X_{1},X_{2}$.\\ 

(ii) With $\pi$ from $X_{1}$, the quantities 
     $\langle \mu \rangle_{\pi}^{*}$ and ${\cal F}_{*}$ are calculated 
     for $X_{2}$ with Eqs. (15)-(17).\\

(iii) $M$ independent data sets
      ${{\cal D}_{m}}$ are now created with 
      $u_{i} = \langle \mu \rangle_{\pi}^{*} + \sigma z_{i}$  .\\

(iv)  For each ${\cal D}_{m}$, the new value of 
      $\chi^{2}(\langle \vec{\alpha} \rangle_{\pi}^{*})$ is calculated 
      for the $\langle \mu \rangle_{\pi}^{*}$ derived at step (ii).\\

(v)   For each ${\cal D}_{m}$, the new value of 
      $\chi^{2}(\langle \vec{\alpha} \rangle_{u})$ is calculated
      with $\langle \vec{\alpha} \rangle_{u}$ from Eq.(16).\\

(vi)  With these new $\chi^{2}$ values, the statistic ${\cal F}_{m}$
      is obtained from Eq.(17).\\

The resulting $M$ values of ${\cal F}_{m}$ give us an empirical estimate
of $p_{*}$, namely $f_{*}$, the 
fraction of the ${\cal F}_{m}$ that exceed ${\cal F}_{*}$.  
In one example of this test, a data pair
$D_{1},D_{2}$ gives $\langle \mu \rangle_{\pi}^{*} = 0.089$ and 
${\cal F}_{*} = 1.0021$. With
$\nu_{1}=1$ and $\nu_{2} = 99$, Eq.(18) then gives $p_{*} = 0.3192$. 
This is checked by creating $M = 10^{5}$
data sets ${\cal D}_{m}$ with the assumption that
$\mu_{true} = 0.089$. The resulting empirical estimate is $f_{*} = 0.3189$,
in close agreement with $p_{*}$ 

From $100$ independent pairs $D_{1},D_{2}$, the mean value of
$|p_{*} - f_{*}|$ is $0.001$. This simulation confirms the accuracy of 
$p$-values derived from Eq.(18) and therefore of decision thresholds 
$F_{\nu_{1},\nu_{2},\beta}$.

\subsection{Bias test}

To investigate this Bayesian $F$-test's ability to detect
inconsistecy between $X_{1}$ and $X_{2}$, the
bias test of Sect.5.2 is repeated, again with $n_{1} = n_{2} = 100$. 
The vector $\vec{\alpha}$ in Sect.6.1 now becomes the scalar $\mu$, 
and $i = 1$. 

In Fig.3, the values of ${\cal F}$ from Eq.(17) with $j = i = 1$ 
are plotted against the bias parameter
$b/\sigma$. Also plotted are critical values
derived from the distribution $F_{\nu_{1},\nu_{2}}$ with $\nu_{1} = 1$ and 
$\nu_{2} = 99$.
In contrast to Fig.2 for the $\langle \chi^{2} \rangle_{\pi}$ statistic,
inconsistency between $X_{1}$ and $X_{2}$ is now detected down to
$b/\sigma \approx 0.5$.

In this test of inconsistency, the flaw in $X_{1}$ is the bias $b$.
Now, if it were known for certain that this
was the flaw, then a Bayesian analysis with $H_{1}$ changed from
$u = \mu$ to $u = \mu +b$ - i.e., with an extra parameter - is
staightforward. The result is the
posterior density for $b$, allowing for correction.
In contrast, the detection of a flaw with $\langle \chi^{2} \rangle_{\pi}$ or 
${\cal F}$ is cause-independent. Although Figs.2 and 3 have $b/\sigma$ as the 
abscissa, for real experiments this quantity is not known and   
conclusions are drawn just from the ordinate.
\begin{figure}
\vspace{8.2cm}
\includegraphics{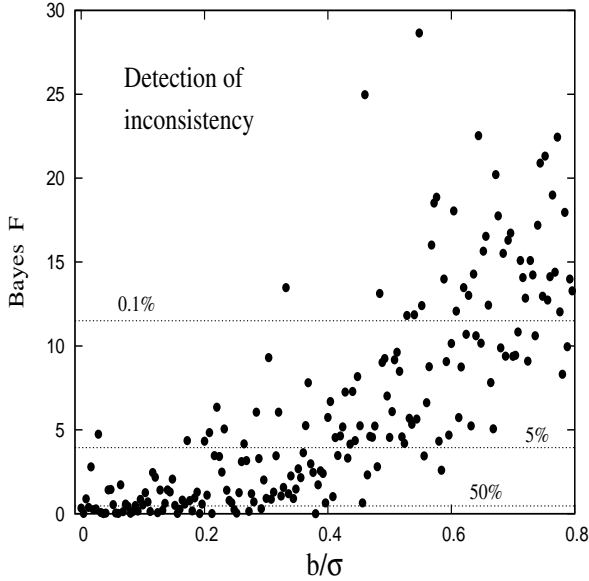}
\caption{Detection of inconsistency between $X_{1}$ and $X_{2}$. 
Values of ${\cal F}$ are plotted against the bias parameter 
$b/\sigma$.
The dashed lines are the $0.1, 5$ and $50 \%$ levels. Note that the
horizontal scale differs from Fig.2} 
\end{figure}
\section{Tension between experiments}
In the above, the goodness-of-fit of  
${\cal M}$ to $D$ is investigated taking into account the possibility that
a poor fit might be due the prior derived from a 
previous experiment. A related goodness-of-fit issue commonly arises in
modern astrophysics, particularly cosmology, namely whether estimates of
non-identical but partially overlapping 
parameter vectors from different experiments are mutually
consistent. The term commonly used in this regard is tension, with 
investigators often reporting 
their {\em subjective} assessments - e.g., there is marginal tension 
between $X_{1}$ and $X_{2}$
- based on the two credibility 
domains (often multi-dimensional) for the parameters in common.

In a recent paper, Seehars et al. (2015) review the attempts 
in the cosmological literature to quantify
the concept of tension, with emphasis on CMB experiments.  
Below, we develop a rather different approach based on the ${\cal F}$
statistic defined in Sect.6.2.

Since detecting and resolving conflicts between experiments is
essential for scientific progress, it is desirable to quantify 
the term tension and to optimize its detection. The conjecture here is 
that this optimum is achieved when inferences from $X_{1}$ are {\em imposed} 
on the analysis of $X_{2}$

\subsection{Identical parameter vectors} 
A special case of assessing tension between experiments is that where the 
parameter vectors are identical. But this is just the problem
investigated in Sects. 5 and 6. 

When $X_{1}$ (with bias) and $X_{2}$ (without bias) are separately
analysed, the limited overlap of the two 
credibility intervals for $\mu$ provides a {\em qualitative} indication 
of tension.
However, if $X_{2}$ is analysed with a prior derived from  $X_{1}$,
then the statistic $\langle \chi^{2} \rangle_{\pi}$ - see Fig.2  - 
or, more powerfully, the statistic ${\cal F}$ - see Fig. 3 - provide  
a {\em quantitative} measure to inform statements about the degree of 
tension.  
\subsection{Non-identical parameter vectors-I} 
We now suppose that $D_{1},D_{2}$ are data sets from different
experiments $X_{1},X_{2}$ designed to test the hypotheses 
$H_{1},H_{2}$. However, though different, these hypotheses have parameters
in common.  
Specifically, the parameter vectors of $H_{1}$ and $H_{2}$ are
$\vec{\xi} = (\vec{\alpha},\vec{\beta})$ and
$\vec{\eta} = (\vec{\beta},\vec{\gamma})$, respectively, and 
$k,l,m$ are the
numbers of elements in $\vec{\alpha},\vec{\beta},\vec{\gamma}$,
respectively.

If $X_{1}$ and $X_{2}$ are analysed independently, we may well find that
${\cal M}_{1}$ and ${\cal M}_{2}$ provide
satisfactory fits to $D_{1}$ and $D_{2}$ and yet still be 
obliged to report tension between the experiments
because of a perceived inadequate overlap of the two $l$-dimensional 
credibility domains for $\vec{\beta}$. 

A quantitative measure of tension between $X_{1}$ and $X_{2}$ can be derived 
via the priors, as follows:
The analysis of $X_{1}$ gives $p(\vec{\xi}| H_{1}, D_{1})$, the posterior  
distribution of $\vec{\xi}$, from which 
we may derive the posterior distribution of $\vec{\beta}$ by integrating
over $\vec{\alpha}$. Thus 
\begin{equation}
 p(\vec{\beta}|H_{1},D_{1}) = \int p(\vec{\xi}| H_{1}, D_{1}) \: 
                                                       d V_{\vec{\alpha}}  
\end{equation}
Now, for a Bayesian analysis of $X_{2}$, we must specify $\pi(\vec{\eta})$
throughout $\vec{\eta}$-space, not just 
$\vec{\beta}$-space . But
\begin{equation}
 \pi(\vec{\eta}) \:= \: \pi(\vec{\beta},\vec{\gamma}) 
                \:   = \: \pi(\vec{\gamma}|\vec{\beta})\: \pi(\vec{\beta})
\end{equation}
Accordingly, what we infer from $X_{1}$ can be {\em imposed} on the analysis
of $X_{2}$ by writing 
\begin{equation}
 \pi(\vec{\eta})  \:   = \: 
         \pi(\vec{\gamma}|\vec{\beta})\: p(\vec{\beta}| H_{1}, D_{1})
\end{equation}
The conditional prior $\pi(\vec{\gamma}|\vec{\beta})$ must now be specified. 
This can be taken to be uniform unless we have prior knowledge
from other sources - i.e., not from $D_{2}$.

With $\pi(\vec{\eta})$ specified, the analysis of $X_{2}$ gives
the posterior density  
$p(\vec{\eta}| H_{2}, D_{2})$. As in Sect.6.2, we now regard this
as a fuzzy constraint on $\vec{\eta}$ from which we compute the sharp
constraint $\langle\vec{\eta}\rangle_{\pi}$ given by Eq.(15). Now,
in general, $\langle\vec{\eta}\rangle_{\pi}$
will be displaced from $\langle\vec{\eta}\rangle_{u}$ given by Eq.(16).
The question
then is: Is the increment in $\chi^{2}(\vec{\eta}|H_{2},D_{2})$ between 
$\langle\vec{\eta}\rangle_{\pi}$ and   
$\langle\vec{\eta}\rangle_{u}$ so large 
that we must 
acknowledge tension between $X_{1}$ and $X_{2}$? 

Following Sect.6, we answer this question by computing
${\cal F}$ from Eq.(17) with $i = j = l+m$,
the total number parameters in $\vec{\eta}$. The 
result is then compared to selected critical values from the
$F_{\nu_{1},\nu_{2}}$ distribution, where $\nu_{1} = l+m$ and
$\nu_{2} = n_{2}-l-m$.
With standard assumptions, ${\cal F}$ 
obeys this distribution if 
$\langle\vec{\eta}\rangle_{\pi}    =  \vec{\eta}_{true}$ - i.e, 
if $\langle\vec{\beta}\rangle_{\pi} =  \vec{\beta}_{true}$ {\em and} 
$\langle\vec{\gamma}\rangle_{\pi} =  \vec{\gamma}_{true}$
- see Sect.6.3.
\subsection{A toy model}

A simple example with one parameter ($\mu$) for $X_{1}$ and two ($\mu,\kappa$)
for $X_{2}$ is
as follows: $H_{1}$ states that $u = \mu$ and 
$H_{2}$ states that $v = \mu + \kappa x$. 
The data set $D_{1}$ comprises $n_{1}$ measurements 
$u_{i} = \mu + \sigma z_{i} + b$, where $b$ is the bias. 
The data set $D_{2}$ comprises $n_{2}$ measurements
$v_{j} = \mu + \kappa x_{j} + \sigma z_{j}$, where the $x_{j}$ are
uniform in $(-1,+1)$. The parameters are $\mu=0, \kappa = 1,
\sigma = 1$ and $n_{1} = n_{2} = 100$. In the notation of Sect.7.2, the
vectors 
$\vec{\beta}, \vec{\gamma}$ contract to the scalars $\mu, \kappa$, whence
$l = m =1$, and 
$\vec{\alpha}$ does not appear, whence $k=0$.

\subsection{Bias test}

In the above, $H_{1}$ are $H_{2}$ are different hypotheses
but have the parameter
$\mu$ in common. If $b = 0$, the analyses of   
$X_{1}$ are $X_{2}$ should give similar credibility intervals for
$\mu$ and therefore no tension. But with sufficiently large $b$, tension
should arise. 

This is investigated following Sect.7.2. 
Applying ${\cal M}_{1}$ to $D_{1}$, we derive
$p(\mu|H_{1},D_{1})$. Then, taking the conditional prior 
$\pi(\kappa|\mu)$ to be constant,
we obtain 
\begin{equation}
 \pi(\mu,\kappa)  \:   \propto  \:  p(\mu| H_{1}, D_{1})
\end{equation}
as the prior for the analysis of $X_{2}$. This gives
us the posterior distribution $p(\mu,\kappa|H_{2},D_{2})$, which is
a fuzzy constraint in $(\mu,\kappa)$-space. Replacing this by the sharp
constraint $(\langle\mu\rangle,\langle\kappa\rangle)$, the constrained 
$\chi^{2}$ is
\begin{equation} 
 \chi^{2}_{c} = \chi^{2}(\langle\mu\rangle,\langle\kappa\rangle| H_{2},D_{2})
\end{equation}
Substitution in Eq.(17) with 
$j = i = 2$, then gives ${\cal F}$ as a measure of the tension between  
$X_{1}$ and $X_{2}$. Under standard assumptions, ${\cal F}$ is distributed
as $F_{\nu_{1},\nu_{2}}$ with $\nu_{1} =2, \nu_{2}= n_{2} -2$ 
if $(\langle\mu\rangle,\langle\kappa\rangle) = 
(\mu,\kappa)_{true}$.

In Fig.4, the values of ${\cal F}$ are plotted against $b/\sigma$ together
with critical values for $F_{\nu_{1},\nu_{2}}$ with $\nu_{1} =2, \nu_{2}= 98$.
This plot shows that tension is
detected for $b/\sigma \ga 0.6$. This is slightly inferior to Fig.3 as
is to be expected because of the more complicated $X_{2}$.
\begin{figure}
\vspace{8.2cm}
\includegraphics{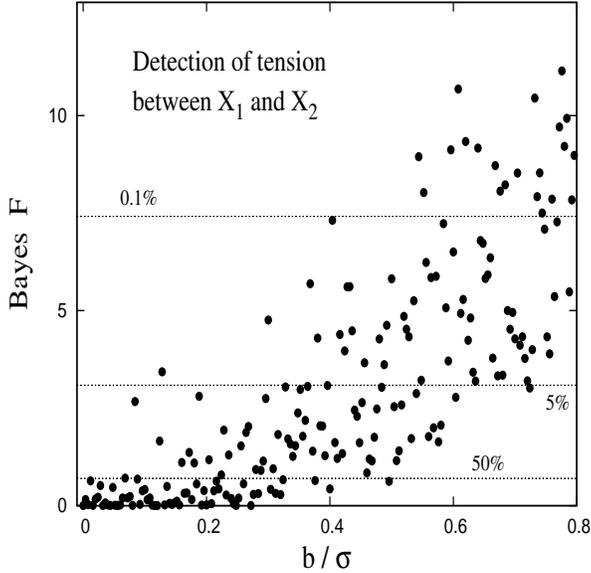}
\caption{Detection of tension between different experiments. Values of 
${\cal F}$ are plotted against the bias parameter $b/\sigma$.
The dashed lines are the $0.1, 5$ and $50 \%$ levels.} 
\end{figure}

 The importance of Fig.4 is in showing that the ${\cal F}$ statistic has the 
desired characteristics of reliably informing the investigator of tension 
between different experiments with partially overlapping
parameter vectors.
When the inconsistency is slight ($b/\sigma \la 0.2$), this statistic does
not sound a false alarm. When the inconsistency is substantial
($b/\sigma \ga 0.6$), the statistic does not fail to sound the alarm.

\subsection{Non-identical parameter vectors-II} 
If ${\cal F}$ calculated as in Sect.7.2 indicates tension, the possible flaws 
include the conditional prior $\pi(\vec{\gamma}|\vec{\beta})$. 
Thus, tension could be indicated even if the  
prior $\pi(\vec{\beta})$ inferred from $X_{1}$ is accurate and
consistent with $D_{2}$.

Accordingly, we might prefer to focus just on $\vec{\beta}$ - i.e., on the
parameters in common. If so, we compute 
\begin{equation}
 \langle\vec{\beta}\rangle_{\pi}  \:   = \: 
         \int \vec{\beta} \: p(\vec{\eta}| H_{2}, D_{2}) \: d V_{\vec{\eta}}
\end{equation}
and then find the minimum of $\chi^{2}(\vec{\eta}|H_{2},D_{2})$ when 
$\vec{\beta} = \langle\vec{\beta}\rangle_{\pi}$.
Thus, the contrained $\chi^{2}$ is now
\begin{equation}
 \chi^{2}_{c}  \:   = \: 
       \min_{\vec{\gamma}} \: \chi^{2}(\langle\vec{\beta}\rangle_{\pi} , \vec{\gamma})
\end{equation}
The $F$ test also applies in this case - see Sect.6.1. Thus this value 
$\chi^{2}_{c}$ is 
substituted
in Eq.(17), where we now take $j = l$, the number of parameters in 
$\vec{\beta}$. The resulting ${\cal F}$ is then compared to critical values
derived from $F_{\nu_{1},\nu_{2}}$ with $\nu_{1} =l, \nu_{2}= n_{2} -l-m$. 
With the standard assumptions, ${\cal F}$ obeys
this distribution if $\langle \vec{\beta}\rangle_{\pi} = \vec{\beta}_{true}$.

For the simple model of Sect.7.3, the resulting plot of ${\cal F}$ against 
$b/\sigma$ is closely similar to Fig.4 and so is omitted. Nevertheless, 
the option of constraining a subset of the
parameters is likely to be a powerful diagnostic tool for
complex, multi-dimensional problems, identifying the parameters
contributing most to the tension revealed when the entire vector is
constrained (cf. Seehars et al. 2015). 

\section{Discussion and Conclusion}
A legitimate question to ask about the statistical analysis of data 
acquired in a scientific experiment is: How well or how badly does
the model fit the data? Asking this question is, after all, just the last 
step in applying the scientific method. In a frequentist analysis, where
the estimated parameter vector $\vec{\alpha}_{0}$ is typically the 
minimum-$\chi^{2}$ point,
this question is answered by reporting the goodness-of-fit statisic
$\chi^{2}_{0} = \chi^{2}(\vec{\alpha}_{0})$ or, equivalently, the corresponding 
$p$-value. If a poor fit is thereby indicated, the investigator and the 
community are aware that not only is the model called into question 
but so also is the estimate $\vec{\alpha}_{0}$ and its confidence domain. 

If the same data is subject to a Bayesian analysis, the same question
must surely be asked: The Bayesian approach does not exempt the investigator 
from the obligation to report on the success or otherwise of the
adopted model. In this case, if the fit is poor, the Bayesian model is
called into question and so also are inferences drawn from the posterior 
distribution $p(\vec{\alpha}|H,D)$.

As noted in Sect.1, the difficulty in testing
Bayesian models is that there are no good fully Bayesian methods for
assessing goodness-of-fit. Accordingly, in this paper, a frequentist
approach is advocated. Specifically, 
$\langle \chi^{2} \rangle_{\pi}$ is proposed in Sect.3 as a suitable 
goodness-of-fit statisic for Bayesian models. Under the null hypothesis
that ${\cal M}$ and $D$ are flawless and with the standard assumptions
of linearity and normally-distributed errors, then, as argued in Sect.4.1
and illustrated in Fig.1,
$\langle \chi^{2} \rangle_{\pi} - k$ is approximately distributed as
$\chi^{2}_{n-k}$, and so $p$-values can be computed. A
$p$-value thus derived from $\langle \chi^{2} \rangle_{\pi}$ quantifies the
average goodness-of-fit provided by the posterior distribution. In contrast,
in a frequentist minimum-$\chi^{2}$ analysis, the $p$-value quantifies the
goodness-of-fit provided by the point estimate $\alpha_{0}$.  

In the above, it is regarded as self-evident that astronomers want to
adhere to the scientific method by always reporting the goodness-of-fit
achieved in Bayesian analyses of observational data. However,
Gelman \& Shalizi (2013), in an essay on the philosophy and practice of
Bayesian statistics, note that investigators who identify Bayesian
inference with inductive inference regularly fit and compare models without
checking them. They deplore this practice. Instead, these authors advocate 
the hypothetico-deductive approach in which model checking is crucial. 
As in this paper, they discuss {\em non-Bayesian} checking of Bayesian models - 
specifically, the derivation of $p$-values from posterior predictive 
distributions. Moreover, they also stress that the prior distribution is a
testable part of a Bayesian model.

In the astronomical literature, the use of frequentist tests to validate 
Bayesian models is not unique to
this paper. Recently, Harrison et al. (2015) have presented an ingenious
procedure for validating multidimensional posterior distributions with
the frequentist Kolmogorov-Smirnov (KS) test for one-dimensional data. Their 
aim, as here, is to test the entire Bayesian inference procedure.

Frequentist testing also arises in recent applications of the 
Kullback-Leibler divergence to quantify tension between cosmological 
probes (e.g. Seehars et al. 2015). For linear models, and with the assumption
of Gaussian priors and likelihoods, a term in the relative entropy is    
a statistic that measures tension. With these assumptions, the statistic
follows a generalized $\chi^{2}$ distribution, thus allowing a $p$-value to 
be computed.

Seehars et al.(2015) also investigate various purely Bayesian measures
of tension. They conclude that interpreting these measures on a fixed,
problem-independent scale - e.g., the Jeffrey's scale - can be misleading
- see also Nesseris \& García-Bellido (2013).

\acknowledgements

I thank D.J.Mortlock for comments on an early draft of this paper, and
A.H.Jaffe, M.P.Hobson and the referee for useful references.

\appendix

\section{Evaluation of $\langle \chi^{2} \rangle_{u}$}

If $\vec{\alpha}_{0}$ denotes the minimum-$\chi^{2}$ solution, then
\begin{equation}
  \vec{\alpha}  =  \vec{\alpha}_{0} + \vec{a} 
\end{equation}
where $\vec{a}$ is the displacement from $\vec{\alpha}_{0}$. Then, on the 
assumption of linearity,
\begin{equation}
  \Delta \chi^{2}(\vec{a})  =  \chi^{2}_{0} + \sum_{i,j}  A_{ij} a_{i} a_{j}  
\end{equation}
where the $A_{ij}$ are the constant elements of the $k \times k$ curvature 
matrix (Press et al. 2007, p.680), where $k$ is the number of parameters. 
It follows that surfaces of constant $\chi^{2}$ are $k$-dimensional 
self-similar ellipsoids centered on $\vec{\alpha}_{0}$.

Now, given the second assumption of normally-distributed measurement errors,
the likelihood
\begin{equation}
   {\cal L} (\vec{\alpha}) \propto \exp 
         \left( -\frac{1}{2} \chi^{2}_{0} \right) 
    \times  \exp \left( -\frac{1}{2} \Delta \chi^{2}   \right) 
\end{equation}
Thus, in the case of a uniform prior, the posterior mean of $\chi^{2}$ is
\begin{equation}
  \langle\chi^{2}\rangle_{u} = \chi^{2}_{0} + 
  \frac{ \int \Delta \chi^{2} \exp ( -\frac{1}{2} \Delta \chi^{2} )  
                                                           dV_{\vec{\alpha}} }
   { \int \exp ( -\frac{1}{2} \Delta \chi^{2} )  dV_{\vec{\alpha}} }
\end{equation}
Because surfaces of constant $\Delta \chi^{2}$ are self-similar,
the $k$-dimensional integrals in Eq.(A.4) reduce to 1-D integrals.

Suppose  $\Delta \chi^{2} =  \Delta \chi^{2}_{*}$ on the surface of the
ellipsoid with volume $V_{*}$. If lengths are increased by the factor 
$\lambda$, then the new ellipsoid has
\begin{equation}
  \Delta \chi^{2} = \Delta \chi^{2}_{*} \times \lambda^{2}   
             \;\; and \;\;   V=V_{*} \times \lambda^{k}
\end{equation}
With these scaling relations, the integrals in Eq.(A.4) can be transformed
into integrals over $\lambda$. The result is
\begin{equation}
  \langle\chi^{2}\rangle_{u} = \chi^{2}_{0} + 2b \:
   \frac{ \int_{0}^{\infty} \lambda^{k+1} \exp ( - b \lambda^{2} )  d\lambda}
   { \int_{0}^{\infty}  \lambda^{k-1} \exp ( - b \lambda^{2} )  d\lambda   }
\end{equation}
where $2 b=\Delta \chi^{2}_{*}$. The integrals
have now been transformed to a known definite integral,
\begin{equation}
   \int_{0}^{\infty} \lambda^{z} \exp ( - b \lambda^{2} ) \: d\lambda
                                = \frac{1}{2} \: \Gamma(x) \: b^{-x}
\end{equation}
where $x = (z+1)/2$. Aplying this formula, we obtain

\begin{equation}
  \langle\chi^{2}\rangle_{u} \: = \: \chi^{2}_{0} + k
\end{equation}
an exact result under the stated assumptions.

\end{document}